# Bilayer Phosphorene: Effect of Stacking Order on Bandgap and its Potential Applications in Thin-Film Solar Cells

*Jun Dai and Xiao Cheng Zeng\**

Department of Chemistry and Department of Mechanical and Materials Engineering, University of Nebraska-Lincoln, Lincoln, NE 68588, USA

**ABSTRACT** Phosphorene, a monolayer of black phosphorus, is promising for nanoelectronic applications not only because it is a natural p-type semiconductor but also it possesses a layer-number dependent direct bandgap (in the range of 0.3 eV~1.5 eV). On basis of the density functional theory calculations, we investigate electronic properties of the bilayer phosphorene with different stacking orders. We find that the direct bandgap of the bilayers can vary from 0.78 - 1.04 eV with three different stacking orders. In addition, a vertical electric field can further reduce the bandgap down to 0.56 eV (at the field strength 0.5 V/Å). More importantly, we find that when a monolayer of $MoS_2$ is superimposed with the p-type AA- or AB-stacked bilayer phosphorene, the combined tri-layer can be an effective solar-cell material with type-II heterojunction alignment. The power conversion efficiency is predicted to be ~18% or 16% with AA- or AB-stacked bilayer phosphorene, higher than reported efficiencies of the state-of-the-art trilayer graphene/transition metal dichalcogenide solar cells.

**TOC GRAPHICS**

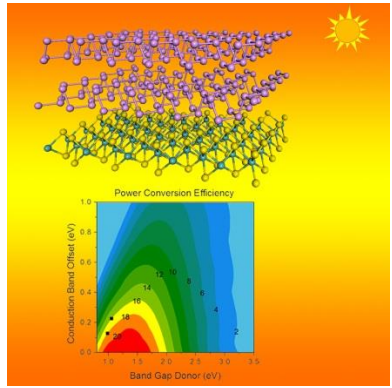

Two-dimensional (2D) materials such as monolayer graphene and monolayer transition metal dichalcogenide (TMDC) MoS$_2$ have attracted intensive research interests owing to their fascinating electronic, mechanical, optical or thermal properties, some of them not seen in their bulk counterparts, *e.g.*, the massless Dirac-fermion behavior of the graphene.[1-3] However, the lack of a bandgap in graphene severely limits its applications in nanotransistors.[4-6] Monolayer MoS$_2$ does possess a direct bandgap of about 1.8 eV.[7] Indeed, the field effect transistor (FET) based on monolayer MoS$_2$ has demonstrated good device performance with a high on/off ratio of ~10$^8$.[8,9] Nevertheless, although the carrier mobility in monolayer MoS$_2$ was previously reported[8] to be approximately 200 cm$^2$/V/s and may be further improved to 500 cm$^2$/V/s,[9] a few recent experiments indicate that the mobility values might be overestimated due to the capacitive coupling between the gates,[10, 11] thereby limiting its wide application in electronics.

Very recently, a new 2D semiconducting material with a direct bandgap, namely, the few-layer black phosphorus (phosphorene), has been successfully isolated.[12-14] Moreover, the phosphorene based FET exhibits high mobility of 286 $cm^2/V/s$ and appreciably high on/off ratios, up to $10^4$.[13] The mobility is thickness dependent and can be as high as ~1000 $cm^2/V/s$ at ~10 nm thickness.[12] Like graphite, black phosphorus, the bulk counterpart of phosphorene, is also a layered material with weak interlayer van der Waals (vdW) interaction. In the monolayer phosphorene, the phosphorus atom is bonded with three adjacent phosphorus atoms, forming a puckered honeycomb structure (see Figure 1). One of novel properties of the phosphorene is the thickness-dependent bandgap. Previous first-principles calculations show that the bandgap ranges from ~1.5 eV for a monolayer to ~0.6 eV for a 5-layer.[15] Bulk black phosphorus however has a direct bandgap of 0.3 eV.[16-18] The bandgap of monolayer phosphorene is also predicted to be highly sensitive to either in-plane or out-of-plane strain. A ~3% in-plane strain can change phosphorene from a direct-gap to an indirect-gap semiconductor,[13] while a vertical compression can induce the semiconductor to metal transition.[19] Since the electronic properties of phosphorene are highly dependent on its thickness, it is of both fundamental and practical interests to attain a better understanding of the effect of interlayer interaction on the electronic properties. Bilayer phosphorene is the thinnest multilayer system that can provide fundamental information on the interlayer interaction and stacking-dependent electronic and optical properties, a feature akin to the bilayer graphene systems.[20-23]

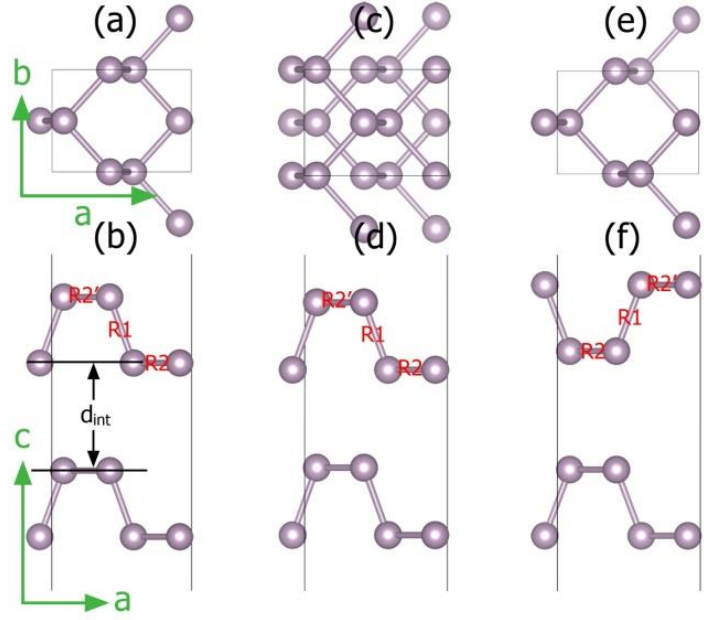

**Figure 1**. Three stacking structures of bilayer phosphorene. (a), (c) and (e) are the top view of AA-, AB- and AC-stacking; (b), (d) and (f) are the side view of AA-, AB- and AC-stacking. R1, R2 and $d_{int}$ denote the two types of P-P bond length and interlayer distance, respectively.

**Table 1**. Optimized lattice constants (*a* and *b*), bond length (R1, R2 and R**2**′) and the nearest distance between the adjacent layer ($d_{int}$) of bilayer phosphorene with three different stacking orders. The unit is Å.

|    | *a*   | *b*   | *R1*  | *R2*  | *R2'* | $d_{int}$ |
|----|-------|-------|-------|-------|-------|-----------|
| AA | 4.550 | 3.326 | 2.283 | 2.243 | 2.235 | 3.495     |
| AB | 4.526 | 3.331 | 2.277 | 2.242 | 2.238 | 3.214     |
| AC | 4.535 | 3.324 | 2.274 | 2.238 | 2.236 | 3.729     |

Three possible stacking orders of bilayer phosphorene are considered, namely, AA-, AB- and AC-stacking. As shown in Figure 1(a), for the AA-stacking, the top layer is directly stacked on the bottom layer. The AB-stacking can be viewed as shifting the bottom layer of the AA-stacking by half of the cell along either *a* or *b* direction (Figure 1(c)). As a result, the edge of the puckered hexagon of the top layer is located in the center of the puckered hexagon of the bottom layer. For the AC-stacking, the top and bottom layers are mirror images of each other. The computed structural parameters are listed in Table 1 where one can see that the lattice constants (*a* and *b*) and bond lengths (R1, R2 and R2') differ slightly for different stacking order, and R2 is always larger than R2'. The most notable difference among the three stacking orders is the nearest distance between the top and bottom layer ($d_{int}$), which varies from 3.214 Å in the AB-stacking to 3.729 Å in the AC-stacking. Our total-energy calculations based on the HSE06 hybrid functional indicate that the AB-stacking is energetically the most favorable, which is 8 meV/atom and 7 meV/atom lower than that of AA- and AC-stacking, respectively.

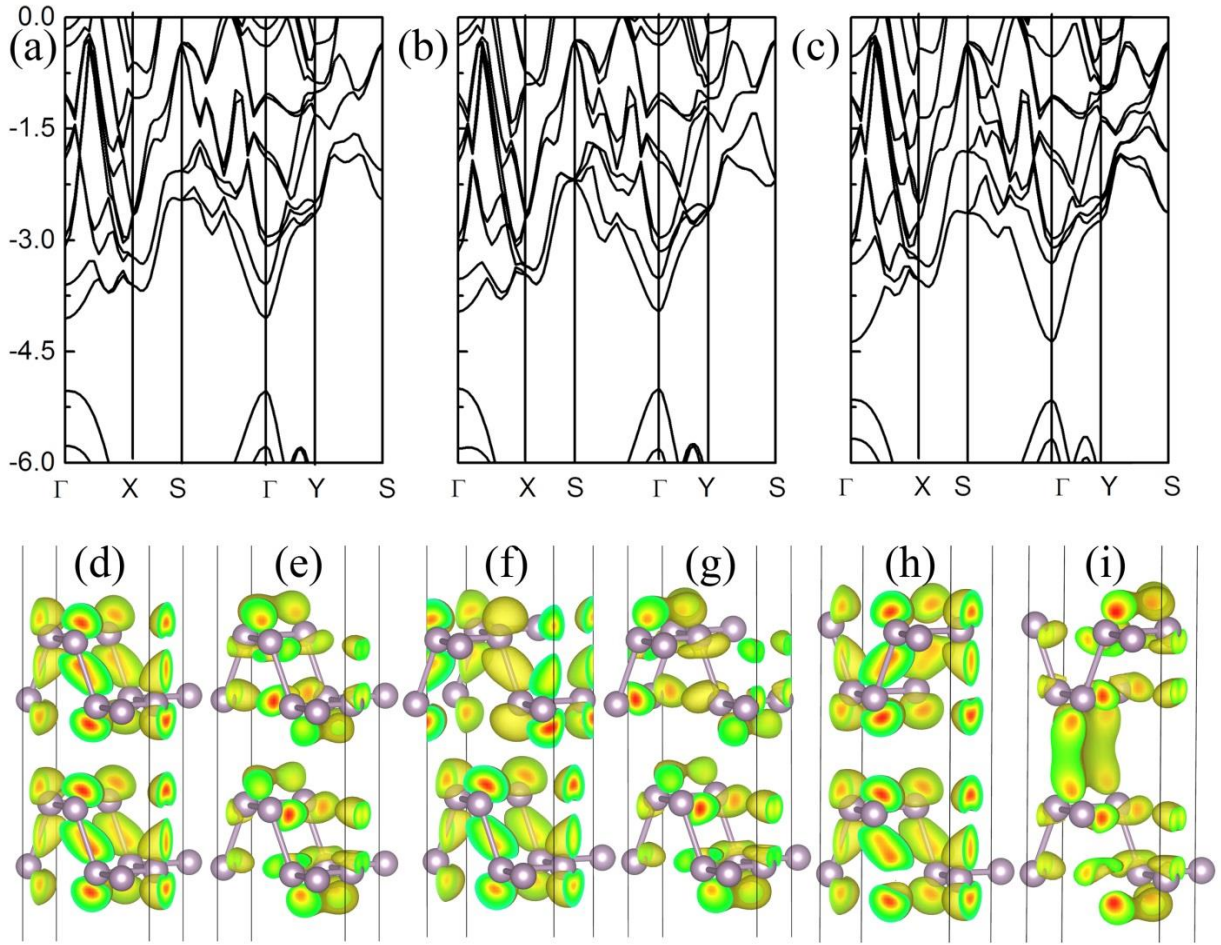

**Figure 2**. Computed band structures (based on HSE06) of (a) AA-, (b) AB- and (c) AC-stacked bilayer phosphorene; (d), (f) and (h) are the iso-surface plot of the charge density corresponding to VBM for AA-, AB- and AC-stacked bilayer phosphorene, while (e), (g) and (i) are the charge density corresponding to the CBM for AA-, AB- and AC-stacked bilayer phosphorene, respectively. The iso-value is 0.004 $e$/Bohr$^3$. Γ (0.0, 0.0, 0.0), X (0.0, 0.5, 0.0), S (0.5, 0.5, 0.0) and Y (0.5, 0.0, 0.0) refer to special points in the first Brillouin zone. The vacuum level is set to be 0 eV.

The HSE06 electronic band structures of AA-, AB- and AC-stacked bilayer phosphorene are shown in Figure 2 (a)-(c). Clearly, the direct-gap feature is retained regardless of the stacking order. Both the valance band maximum (VBM) and conduction band minimum (CBM) are located at the Γ point. Among the three bilayers with different stacking orders, the AB-stacked bilayer possesses the widest bandgap of 1.04 eV, while for the AA- and AC-stacked bilayers, the bandgap is 0.95 eV and 0.78 eV, respectively. By comparing the band structures, we find that the position of VBM with respect to the Fermi level is almost the same, while the CBM of AA- and AC-stacked bilayers is shifted downward, compared to that of AB-stacked bilayer. To understand this difference in the CBM location, we plot the iso-surfaces of the charge density corresponding to VBM and CBM of AA-, AB- and AC-stacked bilayer phosphorene in Figure 2 (d)-(i), respectively. One can see that the VBM is contributed from the localized states of P atoms, while CBM is partly contributed from delocalized states, especially in the interfacial area between the top and bottom layers. Hence, different stacking-order results in different π-π interaction distance between the delocalized states, thereby different interaction strength and bandgap.

Previous theoretical and experimental studies have shown that applying a vertical electric field can tune the bandgap of 2D bilayer graphene and $MoS_2$.[20, 24-29] It is therefore interesting to see how bilayer phosphorene responses to a vertical electric field. The computed bandgap versus the external electric field for bilayer phosphorene with different stacking orders is plotted in Figure 3. Note that within the range of electric field strength considered, the feature of direct bandgap is still retained for all three bilayer phosphorene systems. In general, the bandgap decreases with the increase of the external electric field. The AB-stacked bilayer phosphorene is less sensitive to the external electric field than other two bilayers, and its bandgap decreases from 1.04 eV at 0

V/Å to 0.92 eV at 0.5 V/Å, while the bandgap of AA-stacked bilayer phosphorene is 0.97 eV at 0 V/Å and 0.74 eV at 0.5 V/Å. The AC-stacked bilayer phosphorene is the most sensitive to the electric field and its bandgap changes from 0.78 eV at 0 V/Å to 0.56 eV at 0.5 V/Å. The tunable range of bandgap for the bilayer phosphorene via the vertical electric field is similar to that for the TMDC $MoS_2$/$MoX_2$ heterobilayers.[29,30] All in all, combining the stacking order with electric field, the bandgap of bilayer phosphorene can be tuned in a relatively wide range of 0.56 - 1.04 eV, increasing the tunability for their potential application in nanoelectronics.

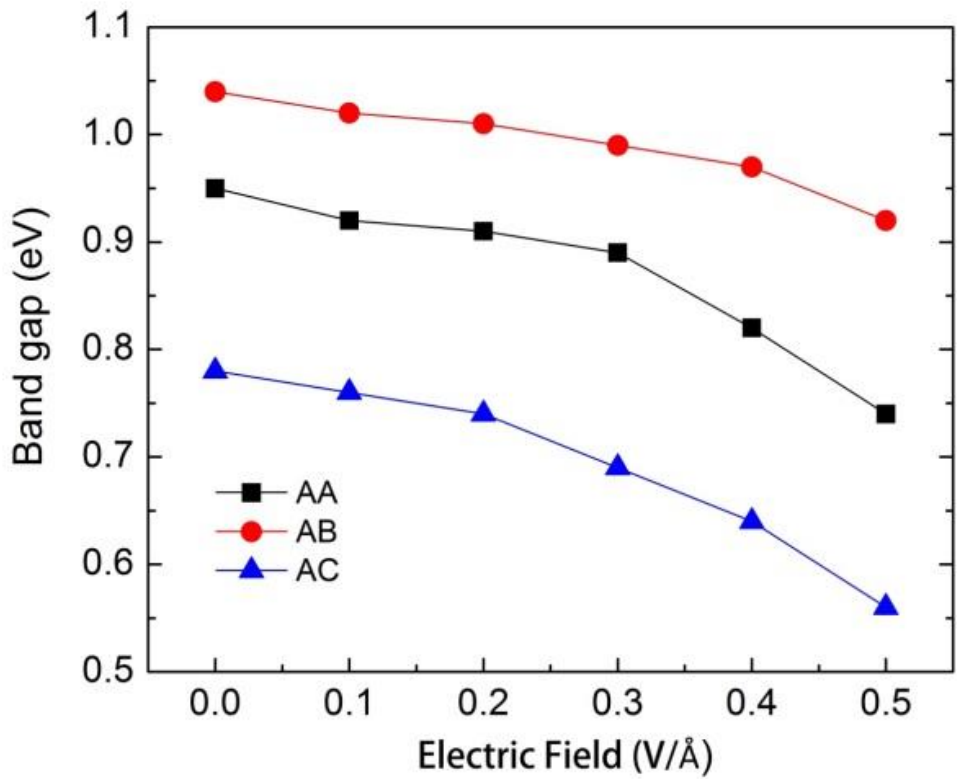

**Figure 3**. Field dependent bandgaps of AA-, AB- and AC-stacked bilayer phosphorene.

The fact that AA-, AB- and AC-stacked bilayer phosphorene possesses direct bandgap of 0.97 eV, 1.04 eV and 0.78 eV, respectively, suggest that the bilayer phosphorene with mixed stacking orders may be very good candidates as solar cell donor materials. To investigate this feasibility, we compute the imaginary part of frequency dependent dielectric function via summing over pairs of occupied and empty states without considering the local field effects, using HSE06 functional. Figure 4(c) shows the computed results of ($\varepsilon_{xx}^2$), one of the diagonal parts of the in-plane componets, for AA-, AB- and AC-stacked bilayer phosphorene. Strong peaks around 1 eV for AA- and AC-stacked bilayers and around 1.4 eV for AB-stacked bilayer can be seen, all arising from the inter-band VBM-CBM transition. There are also peaks due to other inter-band transitions. Overall, the optical absorption is in a relatively wide range from 1 - 4 eV for AA- and AC-stacked bilayers and 1.4 - 4 eV for AB-stacked bilayer, a critical factor to enhance the efficiency of a solar cell. These results confirm that the bilayer phosphorene can potentially serve as donor materials.

From our HSE06 calculations, we find the CBM for AA- -, AB- and AC-stacked bilayer phosphorene are -4.13, -4.03 and -4.43 eV, repectively, and the corresponding VBM are –5.10, -5.07 and -5.20 eV, respectively. If appropiate semiconducting acceptors with matching band alignment can be found, we can build thin-film photovoltaic systems with the bilayer phosphorene. Based on the same (HSE06) method, we find the CBM and VBM of monolayer MoS$_2$ are -4.25 and -6.27 eV, respectively, matching those of AA- and AB-stacked bilayer phosphorene with a type-II heterojunction alignment (Figure 4 (a) and (b)). The upper limit of the power conversion efficiency (PCE) $\eta$ is estimated in the limit of 100% external quantum efficiency (EQE)[31,32] with the formula given by

$$\eta = \frac{J_{sc}V_{oc}\beta_{FF}}{P_{solar}} = \frac{0.65\left(E_g^d - \Delta E_c - 0.3\right)\int_{E_g}^{\infty}\frac{P(\hbar\varpi)}{\hbar\varpi}d(\hbar\varpi)}{\int_0^{\infty}P(\hbar\varpi)d(\hbar\varpi)}$$

where the band-fill factor (FF) is assumed to be 0.65, $P(\hbar\varpi)$ is taken to be the AM1.5 solar energy flux (expressed in W/m$^{-2}$/eV$^{-1}$) at the photon energy $\hbar\varpi$, and $E_g^d$ is the bandgap of the donor, and the $(E_g^d - \Delta E_c - 0.3)$ term is an estimation of the maximum open circuit voltage $V_{oc}$. The integral in the numerator is the short circuit current $J_{sc}$ in the limit of 100% EQE, and the integral in the denominator is the AM1.5 solar flux. As shown in Figure 4(d), solar systems constructed with AA-/AB- stacked bilayer phosphorene and monolayer MoS$_2$ can acheive PCEs as high as ~18% / 16%. These values are comparable to that of the PCBM/CBN system (10-20%)[30] and the recent predicted g-SiC$_2$ based systems (12-20%).[33] Although the proposed application in solar-cell systems is only three-layer thin, thicker multilayers by stacking as the heterojunction could be a viable way to increase the interface area.

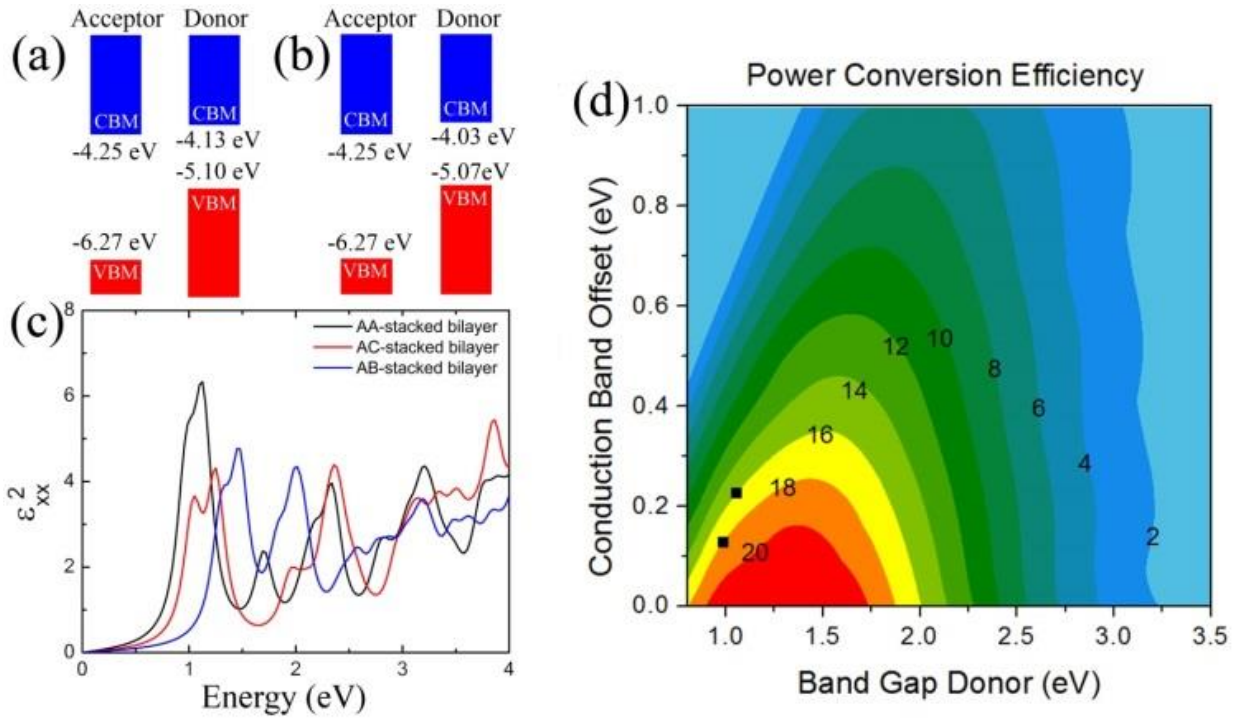

**Figure 4**. (a) Computed band offsets between monolayer MoS$_2$ (acceptor) and AA-stacked bilayer phosphorene (donor); (b) the band offsets between monolayer MoS$_2$ (acceptor) and AB-stacked bilayer phosphorene (donor). The numbers are the CBM and VBM levels with respect to vacuum level. (c) The computed imaginary part of the frequency dependent dielectric function

based on the HSE06 functional. (d) Computed power conversion efficiency contour as a function of the donor bandgap and conduction band offset.

In conclusion, based on DFT calculations, we investigate stacking effect on the electronic and optical properties of bilayer phosphorene. Our calculations show that CBM is very sensitive to the stacking order and can results in a change of the bandgap from 1.04 eV (AB-stacked) to 0.78 eV (AC-stacked). The bandgap also decreases with increasing the vertical electric field, e.g., down to 0.56 eV at 0.5 V/Å field strength. Potential application of mixed bilayer phosphorene as a solar-cell donor material is examined. The predicted PCE for monolayer $MoS_2$/AA-stacked bilayer phosphorene and $MoS_2$/AB-stacked bilayer phosphorene can be as high as ~18% and 16%, respectively, rendering the trilayer $MoS_2$ phosphorene a promising candidate in flexible optoelectronic devices. In view of the recent and successful fabrication of efficient thin-film solar cells using a van der Waals trilayer of graphene/TMDC $WS_2$/graphene by Novoselov and coworkers[40], the trilayer phosphorene and TMDC $MoS_2$ systems may be also fabricated in the near future. We expect the van der Waals trilayer phosphorene/TMDC system is a more efficient solar cell than the trilayer graphene/TMDC systems[34,35] because the former can benefit from the absorption of wider range of wavelength in the solar spectrum, and the type-II heterojunction alignment can allow more efficient hole-electron separation.

**Computational Methods**

In the density functional theory (DFT) calculations, we adopt the generalized gradient approximation (GGA) for the exchange-correlation potential. The plane-wave cutoff energy for wave function is set to 500 eV. The ion-electron interaction is treated with the projected augmented wave (PAW)[36,37] method as implemented in the Vienna *ab-initio* simulation package (VASP 5.3).[38,39] For the geometry optimization, 8×10×4 and 8×10×1 Monkhorst-Pack *k*-meshes are adopted for the bulk phosphorus and bilayer phosphorene, respectively. A vacuum spacing of ~15 Å is used so that the interaction between adjacent bilayers can be neglected. During the geometric optimization, both lattice constants and atomic positions are relaxed until the residual force on atoms are less than 0.01 eV/Å and the total energy change is less than $1.0 \times 10^{-5}$ eV. Previous theoretical calculations have shown that the vdW interaction must be accounted for to properly describe the geometrical properties of black phosphorus.[40] Also, previous theoretical calculations demonstrate that the geometrical and electronic properties of black phosphorus and few layer phosphorene is highly functional dependent.[15,40] The combination of optB88-vdW[41,42] for geometry optimization and HSE06[43] for band structure calculation based on the optB88-vdW structure have been proven reliable for few layer phosphorene systems.[15] Our benchmark calculation on bulk black phosphorus also confirms this combined DFT methods, which gives lattice constants of *a*=4.475 Å, *b*=3.337 Å and *c*=10.734 Å, in good agreement with both the experimental ones (*a*=4.376 Å, *b*=3.314 Å, and *c*=10.478 Å)[44] and previous calculations[15]. Our HSE06 band structure calculation of black phosphorus also gives a direct bandgap of 0.36 eV (see Supporting Information Figure S1), in excellent agreement with the experimental bandgap of 0.3 eV.[16] In VASP, the vertical electric field is treated by adding an artificial dipole sheet in the unit cell.[45] Since the applied vertical electric field can affect the interlayer interaction within the bilayer phosphorene, the electronic structures of the systems are computed based on the fully relaxed geometry of the bilayer phosphorene under the vertical electric field.

ASSOCIATED CONTENT

AUTHOR INFORMATION

Corresponding Author

*xzeng1@unl.edu

Notes

Any additional relevant notes should be placed here.

The authors declare no competing financial interests.

ACKNOWLEDGMENT

This work is supported by ARL (Grant No. W911NF1020099), NSF (Grant No. DMR-0820521), UNL Nebraska Center for Energy Sciences Research and Holland Computing Center, and a grant from USTC for (1000plan) Qianren-B summer research.

**Supporting Information** HSE06 band structures for bulk black phosphorus. This material is available free of charge via the Internet at http://pubs.acs.org.

REFERENCES

(1)    Novoselov, K.; Geim, A. K.; Morozov, S.; Jiang, D.; Grigorieva, M. K. I.; Dubonos, S.; Firsov, A. Two-Dimensional Gas of Massless Dirac Fermions in Graphene. *Nature* **2005,** *438*, 197-200.
(2)    Novoselov, K.; McCann, E.; Morozov, S.; Fal'ko, V. I.; Katsnelson, M.; Zeitler, U.; Jiang, D.; Schedin, F.; Geim, A. Unconventional Quantum Hall Effect and Berry's Phase of $2\pi$ in Bilayer Graphene. *Nature Phys.* **2006,** *2*, 177-180.
(3)    Zhang, Y.; Tan, Y.-W.; Stormer, H. L.; Kim, P. Experimental Observation of the Quantum Hall Effect and Berry's Phase in Graphene. *Nature* **2005,** *438*, 201-204.
(4)    Liao, L.; Lin, Y.-C.; Bao, M.; Cheng, R.; Bai, J.; Liu, Y.; Qu, Y.; Wang, K. L.; Huang, Y.; Duan, X. High-speed Graphene Transistors with a Self-aligned Nanowire Gate. *Nature* **2010,** *467*, 305-308.
(5)    Schwierz, F. Graphene Transistors. *Nature Nanotech.* **2010,** *5*, 487-496.

(6) Wu, Y.; Lin, Y.-m.; Bol, A. A.; Jenkins, K. A.; Xia, F.; Farmer, D. B.; Zhu, Y.; Avouris, P. High-Frequency, Scaled Graphene Transistors on Diamond-like Carbon. *Nature* **2011,** *472*, 74-78.
(7) Mak, K. F.; Lee, C.; Hone, J.; Shan, J.; Heinz, T. F. Atomically Thin $MoS_2$: A New Direct-Gap Semiconductor. *Phys. Rev. Lett.* **2010,** *105*, 136805.
(8) Radisavljevic, B.; Radenovic, A.; Brivio, J.; Giacometti, V.; Kis, A. Single-Layer $MoS_2$ Transistors. *Nature Nanotech.* **2011,** *6*, 147-150.
(9) Yoon, Y.; Ganapathi, K.; Salahuddin, S. How Good Can Monolayer $MoS_2$ Transistors be? *Nano Lett.* **2011,** *11*, 3768-3773.
(10) Fuhrer, M. S.; Hone, J. Measurement of Mobility in Dual-Gated $MoS_2$ Transistors. *Nature Nanotech.* **2013,** *8*, 146-147.
(11) Radisavljevic, B.; Kis, A. Reply to 'Measurement of Mobility in Dual-Gated $MoS_2$ Transistors'. *Nature Nanotech.* **2013,** *8*, 147-148.
(12) Li, L.; Yu, Y.; Ye, G. J.; Ge, Q.; Ou, X.; Wu, H.; Feng, D.; Chen, X. H.; Zhang, Y. Black Phosphorus Field-Effect Transistors. *Nat. Nanotech.* **2014** DOI:10.1038/nnano.2014.35.
(13) Liu, H.; Neal, A. T.; Zhu, Z.; Tomanek, D.; Ye, P. D. Phosphorene: An Unexpected 2D Semiconductor with a High Hole Mobility. *ACS Nano* **2014** DOI: 10.1021/nn501226z.
(14) Reich, E. S. Phosphorene Excites Materials Scientists. *Nature* **2014,** *506* (7486), 19-19.
(15) Qiao, J.; Kong, X.; Hu, Z.-X.; Yang, F.; Ji, W. Few-Layer Black Phosphorus: Emerging Direct Band Gap Semiconductor with High Carrier Mobility. *arXiv preprint arXiv:1401.5045* **2014**.
(16) Warschauer, D. Electrical and Optical Properties of Crystalline Black Phosphorus. *J. Appl. Phys.* **1963,** *34*, 1853-1860.
(17) Asahina, H.; Morita, A. Band Structure and Optical Properties of Black Phosphorus. *J. Phys. C* **1984,** *17*, 1839.
(18) Takahashi, T.; Tokailin, H.; Suzuki, S.; Sagawa, T.; Shirotani, I. Electronic Band Structure of Black Phosphorus Studied by Angle-Resolved Ultraviolet Photoelectron Spectroscopy. *J. Phys. C* **1985,** *18*, 825.
(19) Rodin, A.; Carvalho, A.; Neto, A. Strain-induced Gap Modification in Black Phosphorus. *arXiv preprint arXiv:1401.1801* **2014**.
(20) Zhang, Y.; Tang, T.-T.; Girit, C.; Hao, Z.; Martin, M. C.; Zettl, A.; Crommie, M. F.; Shen, Y. R.; Wang, F. Direct Observation of a Widely Tunable Bandgap in Bilayer Graphene. *Nature* **2009,** *459*, 820-823.
(21) Wang, Y.; Ni, Z.; Liu, L.; Liu, Y.; Cong, C.; Yu, T.; Wang, X.; Shen, D.; Shen, Z. Stacking-dependent Optical Conductivity of Bilayer Graphene. *ACS Nano* **2010,** *4*, 4074-4080.
(22) Kim, Y.; Yun, H.; Nam, S.-G.; Son, M.; Lee, D. S.; Kim, D. C.; Seo, S.; Choi, H. C.; Lee, H.-J.; Lee, S. W. Breakdown of the Interlayer Coherence in Twisted Bilayer graphene. *Phys. Rev. Lett.* **2013,** *110*, 096602.
(23) Zou, X.; Shang, J.; Leaw, J.; Luo, Z.; Luo, L.; Cheng, L.; Cheong, S.; Su, H.; Zhu, J.-X.; Liu, Y. Terahertz Conductivity of Twisted Bilayer Graphene. *Phys. Rev. Lett.* **2013,** *110*, 067401.
(24) Ohta, T.; Bostwick, A.; Seyller, T.; Horn, K.; Rotenberg, E. Controlling the Electronic Structure of Bilayer Graphene. *Science* **2006,** *313*, 951-954.
(25) Castro, E. V.; Novoselov, K.; Morozov, S.; Peres, N.; Dos Santos, J. L.; Nilsson, J.; Guinea, F.; Geim, A.; Neto, A. C. Biased Bilayer Graphene: Semiconductor with a Gap Tunable by the Electric Field Effect. *Phys. Rev. Lett.* **2007,** *99*, 216802.

(26)  Oostinga, J. B.; Heersche, H. B.; Liu, X.; Morpurgo, A. F.; Vandersypen, L. M. Gate-induced Insulating State in Bilayer Graphene Devices. *Nature Mater.* **2007,** *7*, 151-157.
(27)  Mak, K. F.; Lui, C. H.; Shan, J.; Heinz, T. F. Observation of an Electric-Field-Induced Band Gap in Bilayer Graphene by Infrared Spectroscopy. *Phys. Rev. Lett.* **2009,** *102* (25), 256405.
(28)  Ramasubramaniam, A.; Naveh, D.; Towe, E. Tunable Band Gaps in Bilayer Transition-Metal Dichalcogenides. *Phys. Rev. B* **2011,** *84,* 205325.
(29)  Lu, N.; Guo, H.; Lei, L.; Dai, J.; Wang, L.; Mei, W.-N.; Wu, X.; Zeng, X. C. $MoS_2/MX_2$ Heterobilayers: Bandgap Engineering via Tensile Strain or External Electrical Field. *Nanoscale* **2014,** *6*, 2879-2886.
(30)  Lu, N.; Guo, H.; Wang, L.; Wu, X. J.; Zeng, X. C. van der Waals trilayers and superlattices: Modification of electronic structures of $MoS_2$ by intercalation. Nanoscale **2014**, *6*, doi:10.1039/C4NR00783B.
(31)  Bernardi, M.; Palummo, M.; Grossman, J. C. Semiconducting Monolayer Materials as a Tunable Platform for Excitonic Solar Cells. *ACS Nano* **2012,** *6* (11), 10082-10089.
(32)  Scharber, M. C.; Mühlbacher, D.; Koppe, M.; Denk, P.; Waldauf, C.; Heeger, A. J.; Brabec, C. J. Design Rules for Donors in Bulk-heterojunction Solar Cells—Towards 10% Energy-conversion Efficiency. *Adv. Mater.* **2006,** *18*, 789-794.
(33)  Zhou, L.-J.; Zhang, Y.-F.; Wu, L.-M. $SiC_2$ Siligraphene and Nanotubes: Novel Donor Materials in Excitonic Solar Cell. *Nano Lett.* **2013,** *13,* 5431-5436.
(34)  Bernardi, M.; Palummo, M.; Grossman, J. C. Extraordinary Sunlight Absorption and One Nanometer Thick Photovoltaics Using Two-Dimensional Monolayer Materials. *Nano Lett.* **2013,** *13*, 3664-3670.
(35)  Britnell, L.; Ribeiro, R.; Eckmann, A.; Jalil, R.; Belle, B.; Mishchenko, A.; Kim, Y.-J.; Gorbachev, R.; Georgiou, T.; Morozov, S. Strong Light-matter Interactions in Heterostructures of Atomically Thin Films. *Science* **2013,** *340*, 1311-1314.
(36)  Blöchl, P. E. Projector Augmented-wave Method. *Phys. Rev. B* **1994,** *50*, 17953.
(37)  Kresse, G.; Joubert, D. From Ultrasoft Pseudopotentials to the Projector Augmented-wave Method. *Phys. Rev. B* **1999,** *59*, 1758.
(38)  Kresse, G.; Furthmüller, J. Efficient Iterative Schemes for Ab Initio Total-energy Calculations Using a Plane-wave Basis Set. *Phys. Rev. B* **1996,** *54*, 11169.
(39)  Kresse, G.; Furthmüller, J. Efficiency of Ab-initio Total Energy Calculations for Metals and Semiconductors Using a Plane-wave Basis Set. *Comp. Mater. Sci.* **1996,** *6,* 15-50.
(40)  Appalakondaiah, S.; Vaitheeswaran, G.; Lebegue, S.; Christensen, N. E.; Svane, A. Effect of van der Waals Interactions on the Structural and Elastic Properties of Black Phosphorus. *Phys. Rev. B* **2012,** *86*, 035105.
(41)  Dion, M.; Rydberg, H.; Schröder, E.; Langreth, D. C.; Lundqvist, B. I. Van der Waals Density Functional for General Geometries. *Phys. Rev. Lett.* **2004,** *92*, 246401.
(42)  Klimeš, J.; Bowler, D. R.; Michaelides, A. Van der Waals Density Functionals Applied to Solids. *Phys. Rev. B* **2011,** *83*, 195131.
(43)  Heyd, J.; Scuseria, G. E.; Ernzerhof, M. Erratum:"Hybrid Functionals Based on a Screened Coulomb Potential"[J. Chem. Phys. 118, 8207 (2003)]. *J. Chem. Phys.* **2006,** *124*, 219906-219906-1.
(44)  Brown, A.; Rundqvist, S. Refinement of the Crystal Structure of Black Phosphorus. *Acta Crystallographica* **1965,** *19*, 684-685.

(45)     Neugebauer, J.; Scheffler, M. Adsorbate-Substrate and Adsorbate-Adsorbate Interactions of Na and K Adlayers on Al (111). *Phys. Rev. B* **1992,** *46*, 16067.